# High-Performance Monolayer WS$_2$ Field-effect Transistors on High-κ Dielectrics

*Yang Cui, Run Xin, Zhihao Yu, Yiming Pan, Zhun-Yong Ong, Xiaoxu Wei, Junzhuan Wang, Haiyan Nan, Zhenhua Ni, Yun Wu, Tangsheng Chen, Yi Shi\*, Baigeng Wang, Gang Zhang\*,Yong-Wei Zhang and Xinran Wang\**

Yang Cui, Run Xin, Zhihao Yu, Xiaoxu Wei, Junzhuan Wang, Prof. Yi Shi, Prof. Xinran Wang
National Laboratory of Solid State Microstructures School of Electronic Science and Engineering and Collaborative Innovation Center of Advanced Microstructures
Nanjing University, Nanjing 210093, P. R. China
E-mail: xrwang@nju.edu.cn, yshi@nju.edu.cn
Yiming Pan, Prof. Baigeng Wang
National Laboratory of Solid State Microstructures, School of Physics, Nanjing University, Nanjing 210093, P. R. China
Zhun-Yong Ong, Prof. Gang Zhang, Prof. Yong-Wei Zhang
Institute of High Performance Computing, 1 Fusionopolis Way, 138632, Singapore
E-mail: zhangg@ihpc.a-star.edu.sg
Haiyan Nan, Prof. Zhenhua Ni
Department of Physics, Southeast University, Nanjing 211189, China
Yun Wu, Tangsheng Chen
Science and Technology on Monolithic Integrated Circuits and Modules Laboratory, Nanjing Electronic Device Institute, Nanjing, China

The continuous scaling of transistors posts serious challenges in power management in CMOS technology. Two-dimensional (2D) materials, especially semiconducting transition-metal dichalcogenides (TMDs), are considered promising candidates for next generation electronic and optoelectronic applications because of their wide band gap and ultrathin body. [1-6] Among the various TMDs, $MoS_2$ has attracted the most attention. $MoS_2$-based field-effect transistors with high on/off ratio over $10^8$ (Ref. 7), cut-off frequency up to 42GHz,[8] and photodetectors with high sensitivity[9] have been successfully demonstrated. However, one of the factors potentially limiting the use of $MoS_2$ for low-power applications is its relatively low phonon-limited mobility ~200-400$cm^2$/Vs at room temperature.[10,11]

$WS_2$ is another typical semiconducting TMD, with a band gap in the range of 1.3-2.05eV, depending on the number of layers.[12,13] As a result of its low effective mass, the predicted room-temperature phonon-limited electron mobility in monolayer $WS_2$ is over 1000$cm^2$/Vs, which is the highest among semiconducting TMDs.[14] However, the reported experimental electron mobility values to date (up to ~50$cm^2$/Vs at room temperature) are much lower than theoretical predictions.[15-18] Furthermore, the devices exhibit insulating transport behavior at low carrier density. It appears that the charge transport in monolayer $WS_2$ is still dominated by extrinsic factors such as Coulomb impurities (CI), charge traps and defects, similar to the case for $MoS_2$. Therefore, an issue central to the realization of $WS_2$–based device applications is the reduction of these impurities in order to reach intrinsic charge transport. This in turn requires the development of a theoretical framework that links the experimentally observed transport behavior to these external microscopic quantities.

In this work, we report the enhancement of the electron mobility in monolayer WS$_2$ field-effect transistors (FETs) through systematic interface engineering. We compare the mobility in WS$_2$ devices in different configurations of interface modification and find that the density of charge traps can be significantly reduced (by ~49%) by an ultrathin Al$_2$O$_3$ dielectric layer between WS$_2$ and SiO$_2$. The enhancement in electron mobility is even more dramatic when combined with thiol functionalization that further decreases the density of CI. Monolayer WS$_2$ transistors undergone these treatments exhibit a record-high mobility of 83 cm$^2$/Vs (337 cm$^2$/Vs) at room temperature (low temperature), a 2.3 (225) times improvement over the devices on bare SiO$_2$. An empirical model incorporating CI and charge traps is developed to quantitatively fit our experimental data and extract the key microscopic quantities. We find that our model cannot capture the temperature dependence of the mobility at high temperatures, suggesting that other scattering processes such as surface optical (SO) phonon in Al$_2$O$_3$ may play an important role.

Our monolayer WS$_2$ samples were exfoliated from bulk flakes (2D materials CO.). We identified the monolayer samples using atomic force microscopy (AFM, Figure 1b), micro-Raman spectroscopy[19] (Figure 1c) and photoluminescence[12,19,20] (Figure 1d). Backgated FETs were fabricated using standard electron beam lithography with Ti/Pd (20nm/20nm) electrodes. To eliminate the effects of contacts on the mobility measurements, we used exclusively in this work the four-probe structure as shown in Figure 1a. The electrical measurements were carried out in a variable-temperature vacuum probe station after in-situ vacuum annealing at 350K to remove adsorbates and improve contacts.[7]

First, we investigated the effect of the substrate on the electrical transport properties of WS$_2$. To this end, we compared devices on bare 300nm SiO$_2$ substrate and on 10nm $Al_2O_3$ / 300nm $SiO_2$ substrate (insets of Figure 2a and 2b). The 10nm $Al_2O_3$ was grown by atomic layer deposition (ALD), with dielectric constant ε=10 from standard capacitance measurements (see Supporting Information for details). The reason for using the hybrid $Al_2O_3/SiO_2$ substrate instead of $Al_2O_3$ directly on Si is twofold: 1. The gate capacitance of the hybrid substrate is only ~1% smaller than the bare $SiO_2$ substrate, thus facilitating comparison at the same carrier density. This is especially important for analyzing the CI-limited mobility because CI scattering is highly sensitive to free carrier screening and varies with the carrier density.[21-24] 2. The ultrathin layer of $Al_2O_3$ does not introduce extra substrate roughness that could undermine the device performances.[25] Figure S1 show typical AFM images of the $Al_2O_3/SiO_2$ and $SiO_2$ substrates used in this work, with a similar mean square roughness of ~0.2nm.

Figure 2a and 2b show the standard four-probe conductivity $\sigma = \frac{I_{ds}}{\Delta V} \frac{L}{W}$ as a function of the back-gate voltage $V_g$ from 300K to 25K for representative devices on $SiO_2$ (S1) and $Al_2O_3$ (A1) substrate respectively, where $I_{ds}$ is the source-drain current; ΔV, L and W are respectively the voltage difference, distance, and sample width between the voltage probes. At room temperature, the conductivity of A1 exhibits a 200% improvement compared to S1 under the same carrier density $n = C_g V_g = 7.0 \times 10^{12}$ cm$^{-2}$ (the gate capacitance $C_g$ =11.5nFcm$^{-2}$ for $SiO_2$ substrate and 11.4 nFcm$^{-2}$ for $Al_2O_3/SiO_2$ substrate). The field-effect mobility $\mu = \frac{d\sigma}{C_g dV_g}$, on the other hand, shows nearly 100% improvement, reaching 49cm$^2$/Vs for A1 even without prolonged vacuum annealing.[18]

To further investigate the performance improvement on high-κ substrate, we performed variable-temperature electrical measurements. Remarkably, the samples show very different low-temperature behaviors. For S1, σ monotonically decreases during cooling down across the entire range of carrier density studied here, indicating an insulating transport behavior (Figure 2a, 2d). For A1, the transfer curves exhibit a crossover near $V_g \approx 75V$ (corresponding to $n \approx 5.3 \times 10^{12}$ cm$^{-2}$), a signature of metal-insulator transition (MIT). Unambiguously metallic and insulating transport are observed for $n > 6.5 \times 10^{12}$ cm$^{-2}$ and $n < 5.3 \times 10^{12}$ cm$^{-2}$ respectively (Figure 2b, 2e). The temperature dependence of μ also diverges for the two cases (Figure 2c). The mobility of S1 under $n = 7.0 \times 10^{12}$ cm$^{-2}$ monotonically increases as a function of temperature with a highest value of 25cm$^2$V$^{-1}$s$^{-1}$ at 300K. On the other hand, the mobility monotonically decreases with temperature for A1. At $T$=300K (25K), $\mu$=49 cm$^2$V$^{-1}$s$^{-1}$ (140 cm$^2$V$^{-1}$s$^{-1}$), which is 2 (90) times that of S1 under the same temperature.

The transport behavior described above closely resembles that of MoS$_2$ undergone thiol chemical functionalization.[26] Therefore, we adopt the same theoretical model (with slight modifications as discussed below) to analyze the data for WS$_2$. The model involves two important elements: charge traps and CI. The former is responsible for the MIT in MoS$_2$. The density of traps has been shown to be roughly equal to the threshold carrier density of the MIT. The latter is the main limiting factor for mobility at high temperatures, giving rise to the typical power law relationship in the $\mu$-$T$ curve. Compared to the model for MoS$_2$, here we do not include the scattering from phonons and short-range defects. The former is motivated by the fact that theoretical phonon-limited mobility is much higher than current experimental values over the entire temperature range,[14] while the latter is motivated by the absence of experimental evidence of short-range defects in WS$_2$. The details of the calculation and fitting are described in Supporting Information.

With this model, we are able to fit the experimental data (Figure 3c) and extract the density of traps ($n_{tr}$) and CI ($n_{CI}$) (Table S1). The agreement between experiment and theory is excellent considering the simplicity of our model (only two parameters). At low temperatures, the calculated mobility is lower than the experimental data for $SiO_2$ substrate, presumably due to the omission of hopping transport in our model.[26,27] We find that $n_{CI}$ is very similar for device S1 and A1, and the major difference comes from $n_{tr}$ (Table S1). Theoretically, the use of high-κ $Al_2O_3$ can lead to an increase in carrier mobility due to dielectric screening effects (Figure 2f).[21,22] However, the temperature dependence of mobility remains largely unchanged except for the slope at high temperatures. Therefore, the drastic difference between S1 and A1 is mainly due to charge traps, which introduces an exponential term with temperature due to the thermal excitation of carriers to the conduction band.[26] When the carrier density in $WS_2$ is much smaller than $n_{tr}$ (as in the case for S1), the charge traps play a dominant role resulting in the insulating behavior. When the carrier density in $WS_2$ is comparable or larger than $n_{tr}$ (as in the case for A1), the charge traps play minor roles resulting in the metallic behavior. We note that the extracted $n_{tr}$ for device A1 is consistent with the threshold carrier density for MIT, reassuring the validity of our model.

Recently, we found that chemical functionalization of $SiO_2$ using (3-mercaptopropyl) trimethoxysilane (MPS) self-assembled monolayer (SAM) can effectively reduce CI and improve the device performance for $MoS_2$.[26] Here we use the same functionalization scheme to further improve the mobility of $WS_2$ devices on $Al_2O_3$. We characterized the SAM on $Al_2O_3$ by surface roughness and X-ray photoelectron spectroscopy (XPS, Figure S2). The S 2p peak was clearly observed in the XPS spectrum of the MPS-treated substrate. The average mean-square roughness calculated by AFM images was less than 0.2nm, verifying the high quality and uniformity of the SAM. The detailed process of MPS treatment and characterizations are described in Supporting Information.

Figure 3 shows the effect of MPS treatment on $WS_2$ device performances. Under the same carrier density of n≈$1.05\times10^{13}cm^{-2}$ (corresponding to $V_g$=150V), the MPS-treated device (A2, Figure 3b) shows σ=85μS (230μS) at room temperature (low temperature), which is 40% (60%) higher than the as-exfoliated sample on $Al_2O_3$ (A1, Figure 3a). The threshold for the MIT also is slightly lower, indicating the reduction in $n_{tr}$ and disorder. Figure 3c shows the mobility as a function of temperature for the two devices at n=$1.05\times10^{13}cm^{-2}$. The same scaling behavior is observed with the MPS-treated device A2 showing higher mobility over the entire temperature range. This similarity in temperature scaling suggests that the major difference between samples is unlikely to be charge traps, as confirmed by our fitting results (Table S1). For device A2, μ=83$cm^2$/Vs (337$cm^2$/Vs) at room temperature (low temperature), which is, to our best knowledge, the highest experimental value to date. The room temperature (low temperature) mobility is ~2.3 (225) times higher than the device on $SiO_2$ reported here, and 70% (134%) higher than the best values reported by other groups.[17,18] The lines in Figure 3c represent the best fitting results. The main reason for the mobility increase is the ~40% reduction in $n_{CI}$ in device A2 (Table S1). The modeling results agree very well with experiments at low temperature but starts to diverge above 100K, indicating additional scattering sources that are not included in our model. Since the intrinsic phonon-limited mobility is over an order of magnitude higher than the experimental values, we can rule out intrinsic phonons here. Short-range defect is also unlikely since it would introduce a temperature-independent term and cannot explain the discrepancy only at high temperatures. We tentatively assign the most likely source of scattering at high temperature to SO phonons from the dielectric layer. The SO phonons, which are shown to be important for graphene and $MoS_2$ on polar high-k dielectrics,[21,24,28] couple with electrons through the random electric fields created by the dipoles of the oscillating metal-oxide bonds. It is expected that the effect of SO phonons is much stronger on $Al_2O_3$ than $SiO_2$.[29,30] Quantitative modelling of electron scattering by SO phonons will be the subject of future research. We stress that all of the experimental observations are reproducible among different devices. In Figure S4, we show the data for another two devices, in which the key transport properties, including the MIT and scaling of mobility, are reproduced qualitatively and consistently.

Finally, we can quantitatively explain the transport phase diagram in the $WS_2$ devices using our theoretical model. Figure 4 shows the conductivity as a function of temperature and carrier density for device A1 (as-exfoliated) and A2 (MPS-treated) on $Al_2O_3$. The critical points for the MIT are marked by red symbols (also see the solid symbols in Figure 2e). Using the parameters in Table S1, the calculated critical points for MIT (black lines) are in excellent agreement with experiments. The MIT threshold carrier density for different devices is consistent with their $n_{tr}$, as expected from our model. At present, the performance of our monolayer $WS_2$ devices is still limited by CI and charge traps despite the substantial mobility improvement from interface engineering.

In summary, our experimental results and modeling analysis demonstrate that interface engineering is a valuable technique for fabricating high-performance $WS_2$ FETs. We have shown that the combination of a thin layer of $Al_2O_3$ and thiol chemical functionalization significantly improve the mobility of $WS_2$ FETs by reducing the density of charge traps and CI. Our work provides a generic path to improve the device performance of TMD-based FETs.

**Acknowledgements**

Y.C., R.X., and Z.Y. contributed equally to this work. This work was supported in part by National Key Basic Research Program of China 2013CBA01604, 2015CB921600, 2013CB932900; National Natural Science Foundation of China 61325020, 61261160499, 11274154, 61204050; National Science and Technology Major Project 2011ZX02707,Natural Science Foundation of Jiangsu Province BK2012302, BK2011011,BK20130055; Specialized Research Fund for the Doctoral Program of Higher Education 20120091110028, MICM Laboratory Foundation 9140C140105140C14070, and a project funded by the Priority Academic Program Development of Jiangsu Higher Education Institutions.

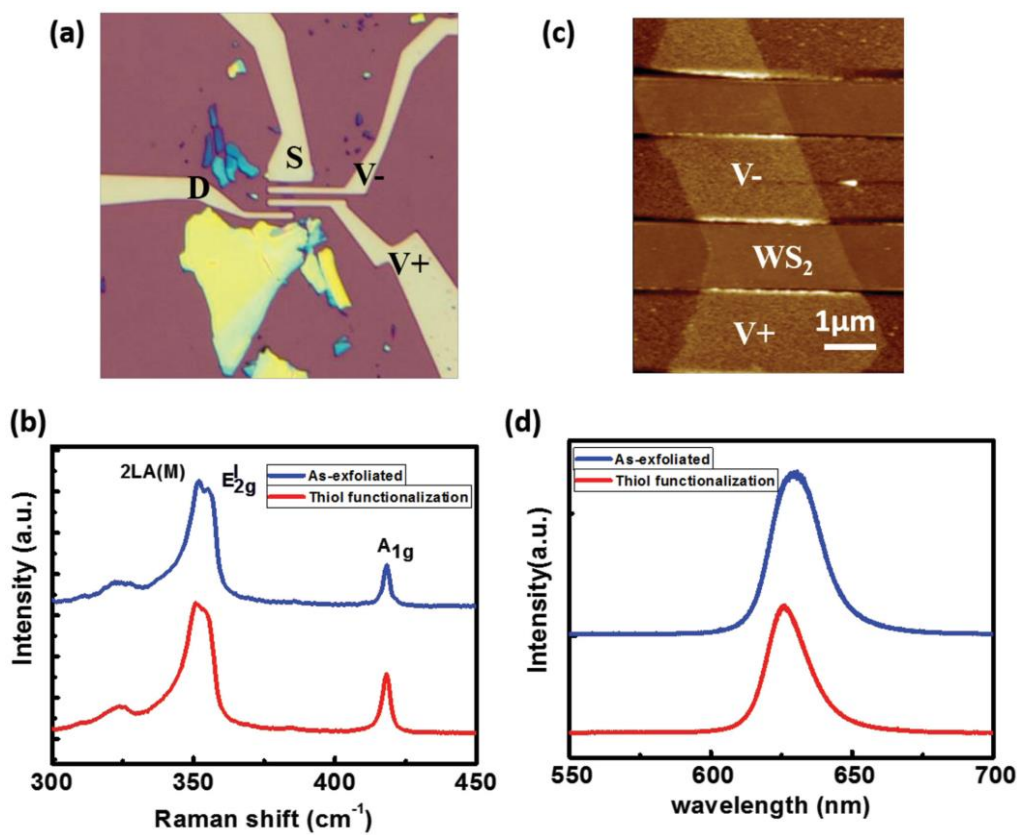

**Figure 1.** Characterizations of monolayer $WS_2$ samples. (a) Optical micrograph of a monolayer $WS_2$ device with four-probe geometry. The source, drain and voltage probes are labeled. (b) AFM image of device in (a). (c) Raman spectrum of a monolayer $WS_2$ as-exfoliated (up) and after MPS functionalization (down) using 514.5 nm laser excitation. The position and relative intensity of the 2LA mode (352 $cm^{-1}$) and the A1 g (417.5 $cm^{-1}$) mode confirm the monolayer thickness. (d) Photoluminescence spectrum of a monolayer $WS_2$ as-exfoliated (blue) and after MPS functionalization (red) using 514.5 nm laser excitation. The Raman and PL do not significantly change after MPS treatment, suggesting that the process does not change the band structure of $WS_2$.

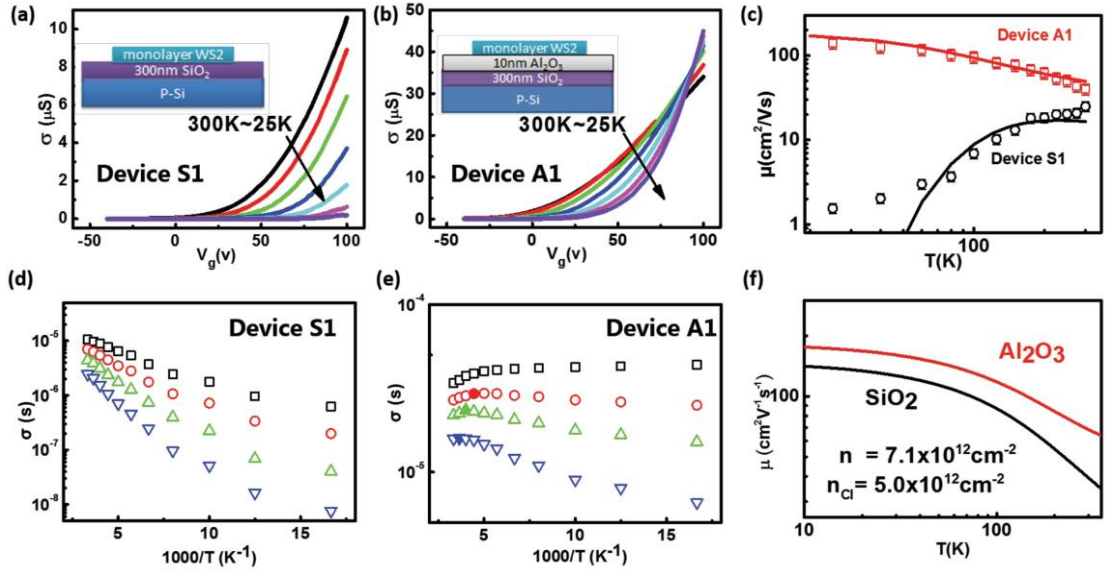

**Figure 2.** The effect of substrate on monolayer WS$_2$ charge transport. (a,b) Typical σ-$V_g$ characteristics for as-exfoliated WS$_2$ samples on (a) 300 nm SiO$_2$/Si and (b)10 nm Al$_2$O$_3$/300 nm SiO$_2$/Si substrate at $T$ = 300, 250, 200, 150, 100, 60, and 25K. Inset: schematics of the devices. (c) $\mu$–$T$ characteristics for the two devices presented in (a) and (b) at $n = 7.1\times10^{12}$ cm$^{-2}$. The error bar is due to the non-rectangular shape of WS$_2$ between the voltage probes. Solid lines represent the best theoretical fitting. (d,e) Arrhenius plot of σ (symbols) and theoretical fittings (lines) for monolayer WS$_2$ on SiO$_2$(d) and Al$_2$O$_3$ (b). From top to bottom $n$ = 7.0, 6.0, 5.0, 4.0$\times10^{12}$ cm$^{-2}$. The solid symbols in (e) are the critical points of MIT. (f) Theoretical CI-limited mobility as a function of temperature for monolayer WS$_2$ on SiO$_2$ and Al$_2$O$_3$ at $n = 7.1\times10^{12}$ cm$^{-2}$ and $n_{CI} = 5.0\times10^{12}$ cm$^{-2}$.

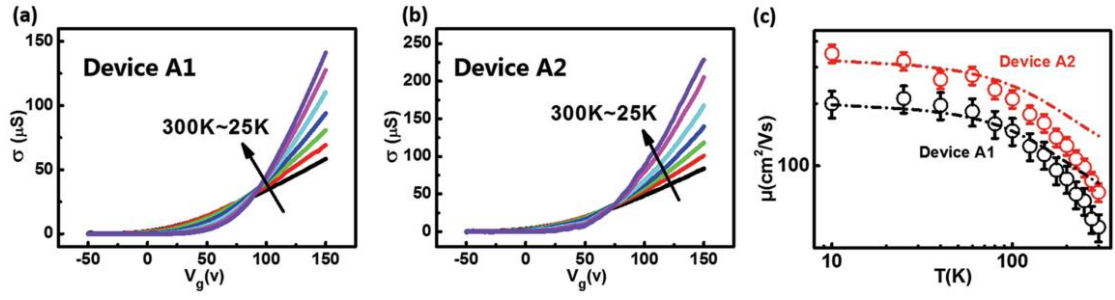

**Figure 3.** The effect of thiol functionalization on electrical transport. (a,b) Typical σ-$V_g$ characteristics for as-exfoliated (a) and MPS-treated (b) monolayer $WS_2$ on $Al_2O_3$ substrate at T = 300, 250, 200, 150, 100, 60, and 25 K. (c) μ-T characteristics for the two devices in (a) and (b) at n = $1.05 \times 10^{13}$ $cm^{-2}$. The error bars are due to the non-rectangular shape of $WS_2$ between the voltage probes. Dash-dotted lines represent the theoretical fitting.

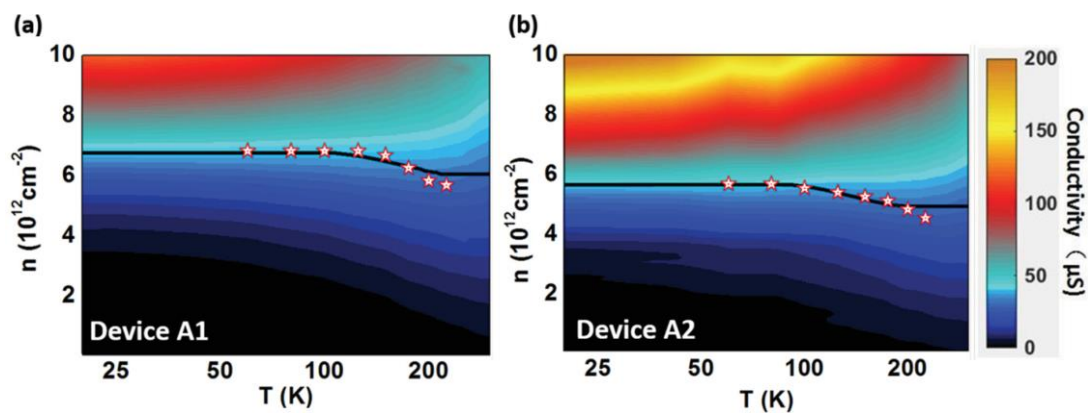

**Figure 4.** Transport phase diagram in monolayer $WS_2$. (a,b) Conductivity as a function of carrier density and temperature for as-exfoliated (a) and MPS-treated (b) monolayer $WS_2$ on $Al_2O_3$ substrate. The solid black lines plot the calculated MIT critical points that separate the metallic and insulating regimes, using the parameters in Table S1 (Supporting Information). The red stars are the experimental critical points of MIT.

# Supporting Information

**S1. Dielectric Screening effect of CI**

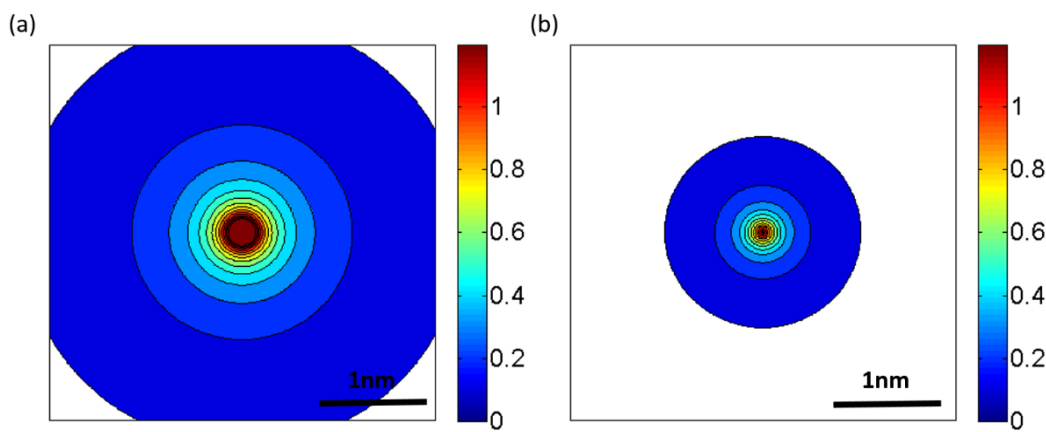

Figure S1. Simulation of the spatial distribution of screened Coulomb potential for a point charge in $MoS_2$ on $SiO_2$ (a) and $HfO_2$ (b). We observe a significant reduction in the effective size of the charge by $Al_2O_3$, due to dielectric screening effects. The effective size of a CI is less than 2nm, which means that 10nm $Al_2O_3$ is thick enough to effectively screen the impurities.

**S2. Growth and characterizations of $Al_2O_3$**

We grew ~10nm thick $Al_2O_3$ on $SiO_2$/Si substrate by Atomic Layer Deposition (ALD). Before ALD, the substrate was cleaned completely by acetone and isopropanol. We adopted trimethylaluminum (TMA, Micro-nano Tech. Co. Ltd., China) and $H_2O$ as precursors to grow $Al_2O_3$ with deposition temperature maintained at 150 °C.

We calculated the average mean-square roughness ($R_q$) of ALD oxide by AFM images. The average $R_q$ of $Al_2O_3$ is ~0.2nm (Fig. S2a), similar to $SiO_2$ (Fig. S2b). This confirms the highly uniform and smooth surface of high-κ oxides, which is important for low charge trap density. The dielectric constant of the oxides was measured by Agilent 4980A precision LCR meter using Au/oxide/Au structure. The capacitance was measured at 10kHz, due to the negligible change of capacitance in low frequency regime.

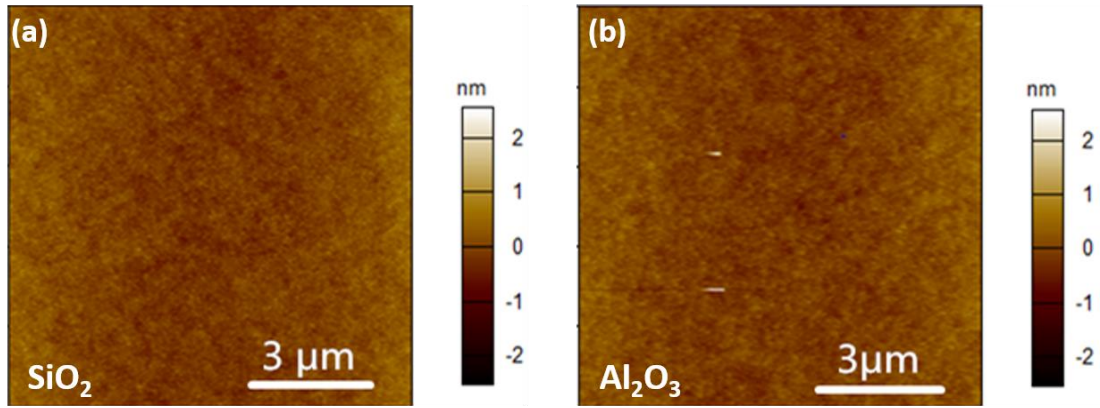

Figure S2. AFM characterization of oxide films (a) bare $SiO_2$, (b) $Al_2O_3$ on $SiO_2$.

## S3. Device fabrication and electrical measurements

We adopted standard electron beam lithography to pattern the electrodes of $WS_2$ devices, followed by electron beam evaporation of Ti/Pd (20nm/20nm) and lift-off. In the ebeam lithography step, we use double layer resist stack (MMA/PMMA), to reduce the exposure dose and form the undercut geometry. We find that such stack does not leave any residue after development, which is better than PMMA alone. Devices were annealed in vacuum at 350℃ for over 30 minutes to improve contacts. Electrical measurements were carried out by a Keithley 4200 semiconductor parameter analyzer in a close-cycle cryogenic probe station with base pressure ~$10^{-5}$ Torr.

## S4. Supplementary electrical data and analysis

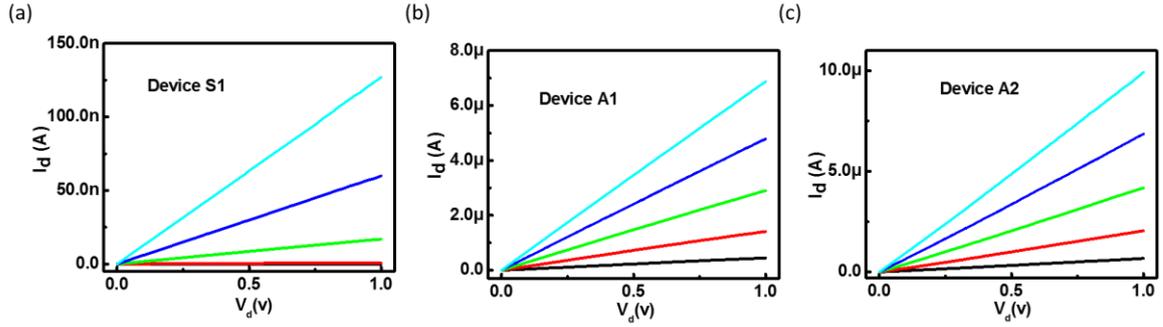

Figure S3. The output characteristics of device S1 (a), A1(b) and A2 (c) at room temperature. From bottom to top, Vg=0V(black), 20V(red), 40V(green), 60V(blue) and 80V(cyan).

In Figure S4 below, we show the typical transfer curve characteristics in log scale for different devices in main text. The room-temperature subthreshold swing (SS) for S1, A1 and A2 are 13.5V/dec, 10.5V/dec and 10V/dec respectively. The interface trap density ($D_{it}$) is related to SS as

$$SS = \ln10\frac{kT}{q}\left(1 + \frac{C_{it}}{Cox}\right) = ln10\frac{kT}{q}(1 + \frac{qD_{it}}{C_{ox}})$$

where $C_{it}$ is the interface trap capacitance, $C_{ox}$ is oxide capacitance, $k$ is the Bolzmann constant, and $T$ is the temperature. We can therefore derive $D_{it}$=1.63×10$^{13}$cm$^{-2}$eV$^{-1}$ (S1), 1.26×10$^{13}$ cm$^{-2}$eV$^{-1}$ (A1), and 1.20×10$^{13}$ cm$^{-2}$eV$^{-1}$ (A2). The derived $D_{it}$ of S1 is very close to that of MoS$_2$ devices on SiO$_2$ (). The $D_{it}$ of A1 and A2 are significantly smaller than S1, which is consistent with our theoretical modeling.

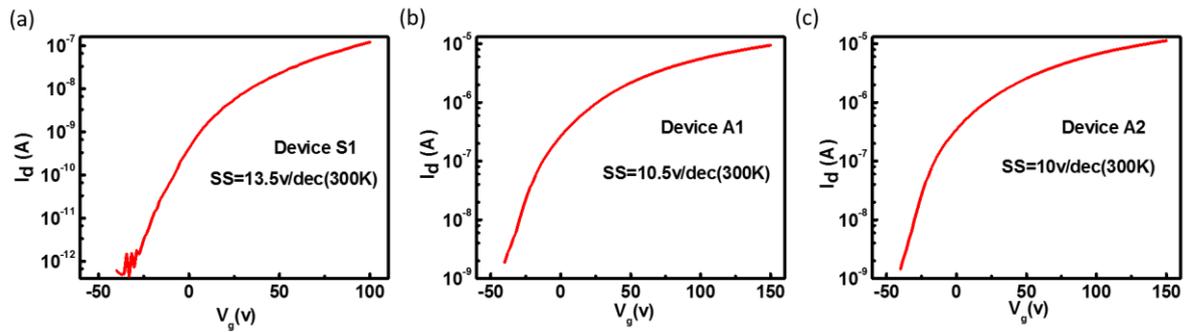

Figure S4. Room-temperature transfer characteristics in log scale of (a) device S1 (b) device A1, and (c) device A2.

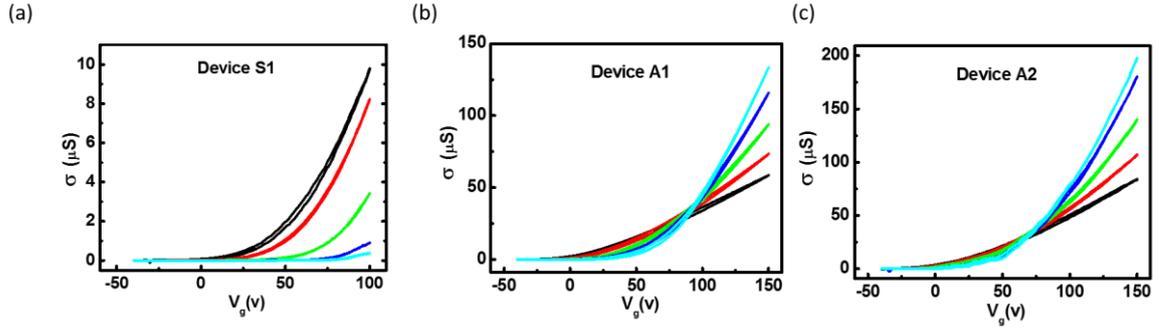

Figure S5. Typical double sweep transfer curve characteristics of (a) device S1 (b)device A1, and (c) device A2. Temperatures: Black: 300K, red: 225K, green: 150K, blue: 80K, cyan: 50K.

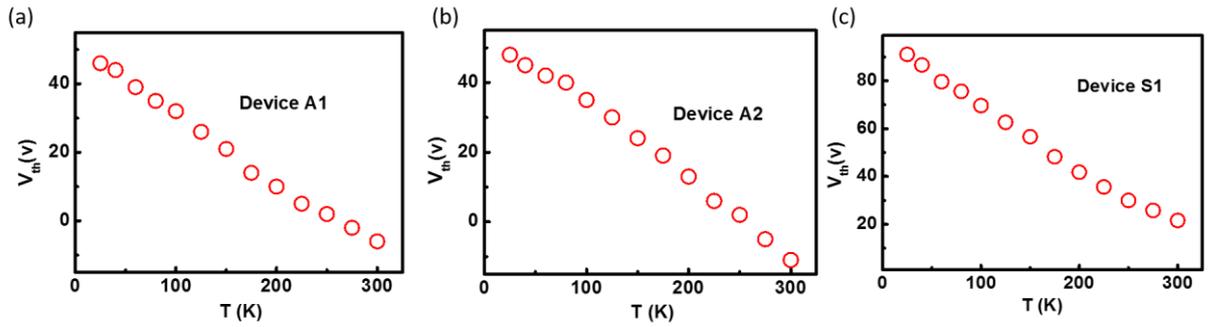

Figure S6. $V_{th}$ as a function of T. From left to right, figure is for device A1 (a), device A2 (b), and device S1 (c).

**S5. Theoretical modeling**

We model the charge carriers in the WS$_2$ as a two-dimensional electron gas with parabolic dispersion and an effective mass of 0.26 $m_0$ (Ref. 1), and assume that the dominant physical mechanism limiting the electron mobility is elastic scattering with charged impurities at the semiconductor-dielectric interface. The CI-limited electron mobility $\mu_{CI}$ is given by the expression[2]

$$\mu_{CI} = \frac{2e}{n\pi\hbar^2 k_B T} \int_0^\infty f(E)[1-f(E)]\Gamma_{CI}(E)^{-1} E\, dE$$

where $e$, $\hbar$, $k_B$, $T$ and $f(E)$ are the electron charge quantum, the Planck constant divided by $2\pi$, the Boltzmann constant, the temperature and the Fermi-Dirac distribution, respectively. The expression for the charged impurity scattering rate $\Gamma_{CI}(E)$ is

$$\Gamma_{CI}(E) = \frac{n_{CI}}{2\pi\hbar} \int_0^\infty d\mathbf{k}' \, |\phi_{|\mathbf{k}-\mathbf{k}'|}^{scr}|^2 \, (1-\cos\theta_{kk'})\delta(E_k - E_{k'})$$

where $\phi_q^{scr}$, $\theta_{kk'}$ and $E_k$ are respectively the screened charged impurity potential, the scattering angle between the $\mathbf{k}$ and $\mathbf{k'}$ states, and the energy of the $\mathbf{k}$ state. Thus, $\mu_{CI}$ is inversely proportional to $n_{CI}$.

The physical effects of the dielectric environment and charge screening of the impurities on the carrier mobility are embedded within the screened CI potential. The expression for the screened CI potential is $\phi_q^{scr} = \frac{\phi_q}{\varepsilon_{2D}(q,T)}$, where $\phi_q = \frac{e^2}{(\varepsilon_{box}^0 + \varepsilon_0)q}$ is the *bare* impurity potential that depends on the dielectric constant of the substrate; $\varepsilon_{box}^0$ and $\varepsilon_0$ are in turn the static permittivity of the substrate and vacuum[2]. The screening of the bare CI by the substrate and the free electrons is described by the generalized screening function $\varepsilon_{2D}(q,T) = 1 + \frac{2\varepsilon_{el}(q)}{\varepsilon_{box}^0 + \varepsilon_0}$, where $\varepsilon_{el}(q) = -\frac{e^2}{2q}\Pi(q,T,E_F)$ corresponds to the electronic part of the dielectric function. $\Pi(q,T,E_F)$ is the temperature- and carrier density-dependent static polarizability, and represents the polarization charge screening of the CI. The exact form of $\Pi(q,T,E_F)$ is given in.[2,3] At high carrier densities, the range of the screened potential is considerably reduced by the polarization charge screening. Thus, the

CI-limited mobility depends on the carrier density and our fitting of the simulated mobility to the one from experiments has to be adjusted for carrier density.

The effect of charge traps is incorporated into the model similar to Ref. 4. We assume the charge traps are uniformly distributed within $\Delta E_{tr}$ below the conduction band edge, with a total density of $n_{tr}$. The Fermi energy $E_F(n, T)$ is determined by

$$n = C_g V_g = \int_0^{+\infty} N_0 \frac{1}{e^{(E-E_F)/k_B T}+1} dE + \int_{-\Delta E_{tr}}^{0} \frac{N_{tr}}{\Delta E_{tr}} \frac{1}{e^{(E-E_F)/k_B T}+1} dE$$

where $N_0 = \frac{2m^*}{\pi \hbar^2} = 2.2 \times 10^{14} \text{eV}^{-1}\text{cm}^{-2}$ is the density of states in the conduction band.[2]

The density of conducting electrons in the extended states is

$$n_c(n,T) = \int_0^{+\infty} N_0 \frac{1}{e^{(E-E_F)/k_B T}+1} dE$$

The conductivity is calculated by

$$\sigma(n,T) = e n_c(n,T) \mu_0(n,T)$$

and the effective mobility is given by

$$\mu(n,T) = \mu_0(n,T) \frac{\partial n_c(n,T)}{\partial n}$$

## S6. Details of MPS treatment

We used double-side MPS chemistry treatment on our $WS_2$ devices on $Al_2O_3$ substrate, similar to Ref. 4. All substrates were subjected to 30-min UV/ozone treatment to hydroxylate the oxide surface. The treated substrates were dipped into 1/10(v/v) MPS/dichloromethane solution for overnight in dry glove box to grow MPS-SAM on the oxide. When MPS growth finished, the substrates were sonicated in dichloromethane followed by thorough rinsing with dichloromethane and IPA, and drying with $N_2$. XPS data in Fig. S7 shows obvious S and Al signal, suggesting that MPS was grown on $Al_2O_3$ substrate successfully. The monolayer $WS_2$ was then exfoliated on the MPS-treated substrate from bulk flakes. Then, the sample was immersed in 1/15(v/v) MPS/dichloromethane solution to grow another layer of MPS on top of $WS_2$. Finally, the sample was annealed in 350℃ for 20 minutes to finish the MPS treatment.

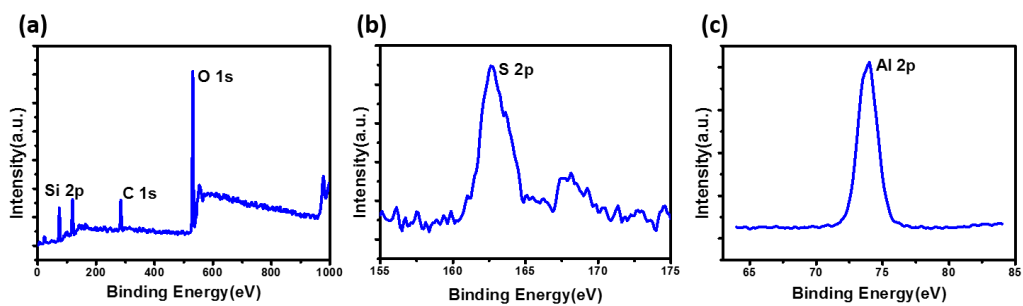

Figure S7. Characterization of MPS grown on $Al_2O_3$ substrate. (a) X-ray photoelectron spectroscopy (XPS) survey scan of MPS SAM grown on $Al_2O_3$ modified $SiO_2$/Si substrate. The signal of Si, O, and C elements are most prominent. (b) High-resolution XPS scan near the S 2p region clearly showing the S signal from MPS. (c) High-resolution XPS scan near the Al 2p region clearly showing the Al signal from $Al_2O_3$.

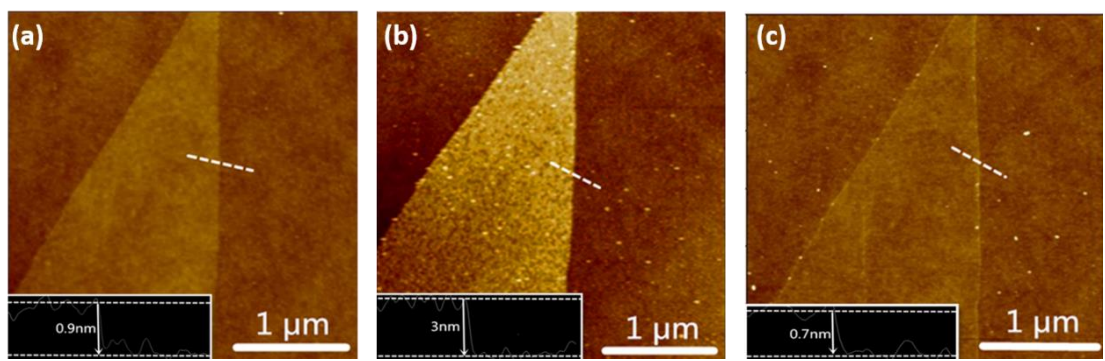

Figure S8. Characterization of $WS_2$ samples subject to MPS treatment. AFM images of the same monolayer $WS_2$ sample (a) exfoliated on $SiO_2$/Si substrate, (b) after immersion in MPS solution and (c) after 350 °C annealing in $H_2$/Ar. The insets show the height of the sample at each stage. The apparent height of $WS_2$ increases significantly after immersion in MPS, indicating that a thick layer of MPS is grown on top of the $WS_2$. After the annealing step, the height of $WS_2$ is restored, indicating that the extra MPS on $WS_2$ is removed.

**S7. Another two sets of transport data for devices on $Al_2O_3$ substrate**

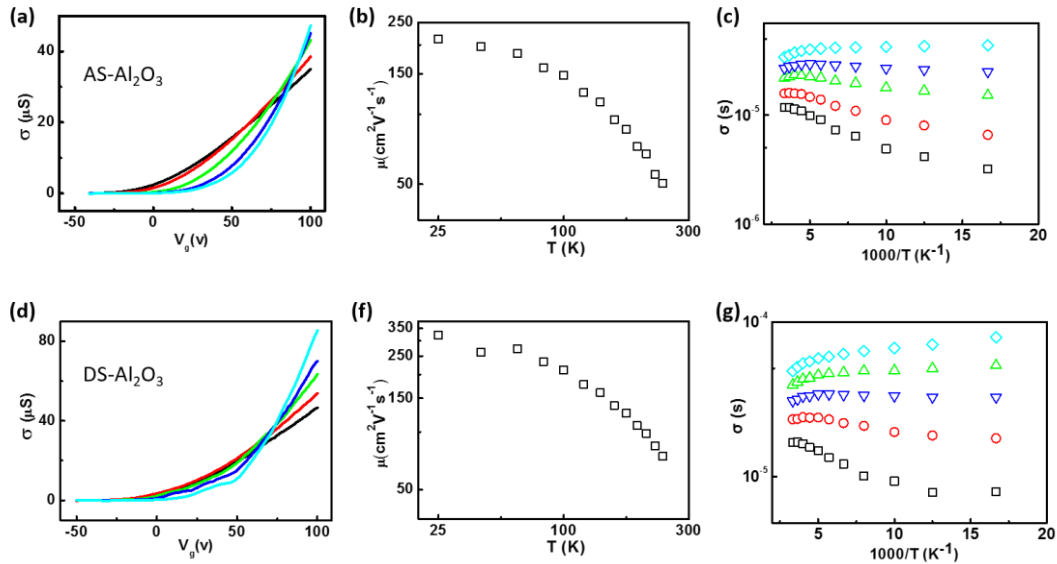

Figure S9. Electrical data of two additional monolayer $WS_2$ devices. (a), (d) are $\sigma$-$V_g$ characteristics for as-exfoliated and MPS-treated monolayer $WS_2$ sample on $Al_2O_3$, respectively. In each panel, black, red, green, blue and cyan curves stand for T=300K, 250K, 150K, 80K, and 25K. (b), (f) are $\mu$-T for the two samples at $n=10.5 \times 10^{12} cm^{-2}$. (c), (g) are Arrhenius plots of $\sigma$ for the two samples. In each panel, from top to bottom, $n= 7.0, 6.0, 5.0, 4.0, 3.0 \times 10^{12} cm^{-2}$ respectively.

**S8. Fitting parameters**

| Device # | SiO$_2$ (ex-foliated, S1) | Al$_2$O$_3$ (ex-foliated, A1) | Al$_2$O$_3$ (MPS-treated, A2) |
|---|---|---|---|
| $n_{CI}$ ($10^{12}$cm$^{-2}$) | 4.5 | 4.9 | 3.0 |
| $n_{tr}$ ($10^{12}$cm$^{-2}$) | 10.1 | 5.13 | 4.4 |
| $\Delta E_{tr}$ (meV) | 84 | 29 | 25 |

Table S1. Fitting parameters of device S1, A1 and A2